# Coherent terahertz radiation with 2.8-octave tunability through chip-scale photomixed microresonator optical parametric oscillation

Wenting Wang[1,†,*], Ping-Keng Lu[2,†], Abhinav Kumar Vinod[1], Deniz Turan[2], James McMillan[1], Hao Liu[1], Mingbin Yu[3,4], Dim-Lee Kwong[4], Mona Jarrahi[2,*], and Chee Wei Wong[1,*]

[1] Fang Lu Mesoscopic Optics and Quantum Electronics Laboratory, University of California, Los Angeles, CA 90095, United States of America

[2] Terahertz Electronics Laboratory, University of California, Los Angeles, CA 90095, United States of America

[3] State Key Laboratory of Functional Materials for Informatics, Shanghai Institute of Microsystem and Information Technology, Shanghai, China

[4] Institute of Microelectronics, A*STAR, Singapore 117865, Singapore

[†] These authors contributed equally to this work.

[*] Email: wentingwang@ucla.edu; mjarrahi@ucla.edu; cheewei.wong@ucla.edu

High-spectral-purity frequency-agile room-temperature sources in the terahertz spectrum are foundational elements for imaging, sensing, metrology, and communications. Here we present a chip-scale optical parametric oscillator based on an integrated nonlinear microresonator that provides broadly tunable single-frequency and multi-frequency oscillators in the terahertz regime. Through optical-to-terahertz down-conversion using a plasmonic nanoantenna array, coherent terahertz radiation spanning 2.8-octaves is achieved from 330 GHz to 2.3 THz, with ≈ 20 GHz cavity-mode-limited frequency tuning step and ≈ 10 MHz intracavity-mode continuous frequency tuning range at each step. By controlling the microresonator intracavity power and pump-resonance detuning, tunable multi-frequency terahertz oscillators are also realized. Furthermore, by stabilizing the microresonator pump power and wavelength, sub-100 Hz linewidth of the terahertz radiation with $10^{-15}$ residual frequency instability is demonstrated. The room-temperature generation of both single-frequency, frequency-agile terahertz radiation and multi-frequency terahertz oscillators in the chip-scale platform offers unique capabilities in metrology, sensing, imaging and communications.

**Introduction**

Terahertz radiation, typically referred to frequencies from 300 GHz to 10 THz ($\lambda \approx$ 30-1000 µm), has spurred remarkable advances in condensed matter physics, biology and medical sciences, global environmental monitoring, metrology, information and communications technology. To achieve high frequency resolution (MHz or lower) and broad bandwidth (100 GHz or more) requirements simultaneously for frequency-agile imaging, communications, metrology, and spectroscopy applications, both widely tunable continuous-wave (CW) terahertz sources as well as terahertz frequency comb sources are demanded extensively. The current frontier techniques for the generation of tunable CW terahertz radiation include millimetre-wave oscillator frequency multiplication [1-2], dual infrared wavelengths photomixing [3-6], as well as using quantum-cascade lasers (QCLs) [7-9], molecular gas lasers [10], and free-electron lasers [11]. Frequency-multiplication approaches can offer milliwatt level radiation power, but hardly reach beyond 1 THz in frequency. In addition, their frequency bandwidths are often limited due to impedance matching and/or waveguide cutoff constraints.

Photonics paves the way to the realization of reliable high-frequency terahertz radiation sources. Notably, much recent progress is offered by the development of terahertz QCLs based on inter-miniband transitions within the conduction band of semiconductor heterostructures. Several realizations of tunable QCLs have been recently demonstrated by intracavity difference-frequency generation [12, 13], microelectromechanical-based transverse mode manipulation [14], and metasurface-based vertical-external-cavity surface emission [15]. Nevertheless, their practical application is often limited by low duty cycle (quasi-CW) operation, restricted frequency tuning range, and the need for cryogenic cooling. Aside from QCLs, a widely tunable terahertz gas laser based on discrete rotational transitions of molecular gas was recently demonstrated with CW operation [10]. However, apart from a relatively larger system footprint due to the required vacuum pump and pressure control elements, its emission frequencies have not surpassed 1 THz so far. Photomixing, which is by far the most reliable technique to achieve wide frequency tunability, CW generation and room temperature operation, is the process of converting a pair of beating optical waves into terahertz radiation at their frequency difference [16]. Unlike nonlinear difference-frequency generation, photomixing in a photoconductor does not require phase matching and its conversion efficiency is not constrained by the Manley-Rowe limit since the terahertz radiation stems from the acceleration of electron-hole pairs rather than direct photon conversion [17, 18]. Therefore, it has been widely recognized in state-of-the-art terahertz generation for frequency-domain spectroscopy [19, 20] and communication systems [21, 22]. Furthermore, terahertz frequency combs have been explored for many high-resolution, high-accuracy metrology and spectroscopy applications, and the typical techniques for the generation of terahertz frequency combs include direct synthesis within QCLs [23], dispersion compensation of QCLs [24], down-conversion and frequency division from stabilized femtosecond mode-locked lasers via photomixing [25, 26]. Besides the QCL and mode-locked laser, a semiconductor injection laser with three different active regions is designed to obtain octave-spanning terahertz frequency combs [27]. The demonstrated terahertz frequency combs can be challenging in terms of the system complexity, small repetition rate, and sizable footprint for cryogenic cooling, limiting some applications outside the laboratory. In parallel, microresonator-based nonlinear processes are proven to provide coherent parametric oscillation generation [28-30], octave-spanning spectral translation [31, 32], and precision broadband frequency combs [33-36] via cavity-enhanced four-wave mixing which are advancing research in laser spectroscopy [37], optical coherent communications [38], optical neural networks [39, 40], laser ranging [41-44], and optical clocks [45, 46]. High-power and coherent millimeter waves are generated recently with the microresonators combining with optical frequency division, dispersion compensation, and optical spectral shaping techniques [47, 48], but the frequency tuning range and resolution can be limited.

Here we present a coherent frequency-agile terahertz wave synthesizer with 2.8-octave tunability, sub-Hz linewidths, and $10^{-15}$ 1,000-sec relative frequency instabilities, capable of generating both single-frequency and multi-frequency terahertz oscillators, through a microresonator-based optical parametric oscillator integrated with a plasmonic nanoantenna array photomixer. First, multi-segmented cavity mode dispersion engineering and polarized hybrid mode interactions in the nonlinear microresonator enables the broadband $\chi^{(3)}$ optical parametric oscillation over 2.3 THz, including competition between mode-crossing Turing instabilities and Faraday instabilities. Secondly, with plasmonic photomixed optical-to-terahertz frequency division, our chip-scale platform achieves the coherent single-frequency synthesis from 330 GHz to 2.3 THz, the largest range realized to our knowledge, and with fine ≈ 27 Hz tuning steps on the terahertz carrier. The generated coherent signal has a cavity-mode-limited frequency tuning step of ≈ 20 GHz and ≈ 10 MHz intracavity mode continuous frequency tuning range at each step. Our terahertz synthesizer does not require optical spectral shaping and, via control of the microresonator intracavity power, multi-frequency terahertz oscillators are also observed through heterodyne beat between our pump laser and the localized optical sub-comb lines. Third, with feedback stabilization of the microresonator intracavity power, we observed less than 100 Hz linewidths on the terahertz radiation, bounded only by the instrument limits. With dual-stage pump power- and frequency-stabilization, the long-term frequency stability is improved by five orders-of-magnitude, reaching an instrument-limited frequency residual instability of $1.6 \times 10^{-15}$ at 1,000 seconds averaging. The chip-scale 2.8-octave tunable synthesizer, with coherent sub-Hz linewidths and $1.6 \times 10^{-15}$ frequency instabilities, at room temperature provides an alternative platform for next-generation terahertz imaging, sensing, metrology and communications.

**Results**

**Chip-scale frequency-agile terahertz radiation synthesizer based on a nitride microresonator and a InAs nanoantenna array.** A conceptual schematic of the broadly-tunable terahertz radiation synthesizer based on a dispersion-managed silicon nitride nonlinear microresonator and a plasmonic photoconductive InAs nanoantenna array photomixer is illustrated in Figure 1a. The tunable optical parametric oscillation (OPO) generation setup is shown in inset **i** and potential applications of the tunable terahertz synthesizer are illustrated in inset **ii**. The microresonator contains seven tapered waveguide segments to engineer the dispersion with varying widths from 1 to 2.5 µm, five 180° bends and two 90° bends for a fixed silicon nitride height of 800 nm. The curved waveguides serve as mode-filters [49, 50] to maintain the fundamental transverse mode ($TE_{00}$ and $TM_{00}$) operation. The averaged cavity group velocity dispersion (GVD) of the fundamental TE and TM modes are 18.31 and 158.10 fs$^2$/mm extracted from the cold-cavity mode spectra measured by swept-wavelength interferometry [36] after the mode frequency calibration (detailed in Methods and Supplementary Information I). The loaded quality factors are $1.55 \times 10^6$ ($TE_{00}$) and $1.23 \times 10^6$ ($TM_{00}$) around the pump laser mode, with mode free spectral ranges (FSR) of 19.87 and 19.72 GHz respectively. The resonant frequencies of the fundamental mode family are $\omega(\mu) = \omega_0 + D_1\mu + \frac{D_2}{2}\mu^2$ when just considering GVD, where $\omega_0/(2\pi)$ is the pump cavity mode frequency, $D_1/2\pi$ is the cavity mode FSR, $\mu$ is the azimuthal mode number with respect to the pump cavity mode, $D_2 = -c/n_0 D_1^2 \beta_2$ is the cavity mode dispersion, $c$ is the speed of the light, $n_0$ is the refractive index of the nitride resonator, and $\beta_2$ is the GVD. To effectively compensate the phase mismatch $\beta_2 \Omega^2$ ($\Omega$ is the parametric oscillating frequency spacing) and $\chi^{(3)}$ Kerr nonlinear phase shift $\gamma P_{in}$ ($\gamma$ is nonlinear coefficient and $P_{in}$ is microresonator intracavity power) in our

normal-dispersion microresonator, mode-frequency shift induced by the transverse mode interaction between our hybridized polarization orthogonal modes (TE and TM) is utilized. Figure 1b shows the computed mode interaction evolution based on the operation temperature by solving the coupled mode equations (detailed in Supplementary Information I). The target cavity mode $\omega_\mu$ is split into two modes $\omega_\mu \pm \Delta\omega$, where $\Delta\omega/(2\pi) \approx \mu(D_{1TE}- D_{1TM})/2\pi + \mu^2(D_{2TE}- D_{2TM})/4\pi \approx \mu c/L(T)/(n_{TE}(T)- n_{TM}(T))/2\pi$ is the mode-shift frequency, $n_{TE}(T)$ and $n_{TM}(T)$ are the temperature-dependent refractive indexes for TE and TM modes, and $L(T)$ is the physical cavity length. Insets **i**, **ii**, and **iii** show the representative mode spectra at different chip temperatures, fitted to the double-Lorentzian model to obtain the mode linewidth and mode-splitting frequency.

Three modes ($\omega_0$, $\omega_{+\mu}$, $\omega_{-\mu}$) are involved in the excitation of the parametric oscillation. By directly modifying the frequency of the target mode through mode-shifting, the tunable phase matching condition is obtained over a broad optical spectral range via $2\omega_0-\omega_\mu-\omega_{-\mu} = \beta_2\Omega^2+\gamma P_{in}- \varphi(\omega_\mu, T)$, where $\varphi(\omega_\mu, T) = \beta_1(\Delta\omega/2\pi) + \beta_2\Omega\Delta\omega/(2\pi)$ and $\beta_1$ is the group velocity. The required mode-splitting frequency is $\Delta\omega/2\pi = (\beta_2\Omega^2+\gamma P_{in})/(\beta_1+\beta_2\Omega)$. The target mode frequency and the corresponding splitting frequency both depend on the microresonator temperature, which is related to the intracavity power $P_{in}(\delta)$ based on the pump-resonance detuning $\delta = (\omega_p-\omega_0)/(2\pi)$. Figure 1c shows the measured tunable parametric oscillation generation via continuous pump-resonance detuning of $\delta = 4$ GHz when the resonant mode is pumped at 1588.15 nm. The resultant parametric signal sidebands tunes from 1585.4 to 1571.5 nm, corresponding to a $\Omega$ tuning from 370 GHz to 1.7 THz ($\Delta\Omega = 1.33$ THz). This specific microresonator thus offers an optical-to-terahertz tuning ratio $\chi = \Delta\Omega/\delta$ of 332.5.

Substantiating our measurements shown in Figure 1, the $\chi^{(3)}$ phase matched parametric sideband curves are calculated based on the modified phase matching condition $L\frac{\beta_2}{2}\Omega^2 + 2\gamma L P_{in}(\delta)^2 - \delta - \varphi(\omega_\mu, T) = 0$ after considering the cavity boundary condition in the Lugiato-Lefever (LL) model and the thermal-mediated mode-splitting as shown in Figure 1c with orange (idler) and purple (signal) curves. The calculated parametric sidebands are consistent with our measurements. Mode-splitting is a local effect which breaks the phase-matching symmetry of the signal ($\omega_{+\mu}$) and idler ($\omega_{-\mu}$) leading to asymmetric parametric gain – Figure 1d shows the resultant measured and calculated signal and idler peak powers. The power difference is attributed to the asymmetric parametric gain, calculated as $G = \sqrt{(\gamma P_{in})^2 - \Delta k}$, where $\Delta k$ is phase mismatch. The computed parametric gain profiles are also shown in Figure 1d where the blue curve is the calculated parametric gain profile with anomalous dispersion as a reference, and $\sigma = \delta + \varphi$ represents the extra phase mismatch in the parametric gain symmetry breaking. After adding the extra mode frequency shift $\Delta\omega/(2\pi)$, the parametric oscillation is generated at effective pump-resonance blue detuning. Figure 1e shows the representative tunable parametric oscillation at different normalized pump-resonance detunings [$\delta_0 = \delta/D_{1TE}/(2\pi)$], where the power imbalance between signal and idler from 10.6 dB ($\delta_0 = -0.4841$) to 0.8 dB ($\delta_0 = -0.0383$) originates from the additional mode-frequency shift. The pump-to-signal power conversion efficiency is related to the pump-to-signal power imbalance ratio (denoted in dB in Figure 1e) determined by the pump resonance detuning and avoided mode frequency shift. A higher pump-to-signal conversion (smaller power imbalance in dB) allows for a higher terahertz wave emission power.

**Thermal and local dispersion dependences of the tunable terahertz optical parametric oscillation.** Tunability of the parametric oscillation is associated with the differential TE-TM thermo-optic effects in the microresonator, demonstrated by monitoring the parametric oscillation at different operating temperatures from 51.74 to 53.07 °C for a fixed pump laser power and the same resonant mode frequency, as shown in Figure 2a. The frequency of the initial parametric oscillation pair can be tuned over 12 nm when the temperature changes 1.3 °C. The power variation of the parametric sidebands is related to the magnitude of the $\Delta\omega/(2\pi)$. Maximizing the spectral tuning range of the parametric oscillation requires a careful selection of the pump cavity mode. By pumping the cavity modes located between 1582 to 1590 nm with a 1 or 2 nm spectral step at a fixed on-chip pump power, we examine the parametric oscillation generation dynamics. At the shorter pump wavelengths, the tunable parametric oscillation vanishes but Faraday parametric oscillation [51] emerges due to the periodic dispersion design in our microresonator [49]. When resonant modes at longer wavelengths are pumped, bistable switching dynamic between Faraday and Turing parametric oscillation is observed, which results from parametric gain competition between the two parametric oscillation processes [52]. The Faraday branch is induced by periodic modulation of the microresonator dispersion. The gain peak sideband frequency away from pump laser frequency is $\omega_f = \sqrt{\dfrac{2\left[\left(\dfrac{\delta}{L}-2\gamma P_{in}\right)\pm\sqrt{(\gamma P_{in})^2+\left(\dfrac{\pi}{L}\right)^2}\right]}{\beta_2}}$, where $L$ is microresonator cavity length. To thoroughly denote the pump wavelength dependence, the high frequency parametric sidebands (signal) are filtered out to highlight the frequency tuning range and parametric oscillation competition dependence on the pump-resonance detuning as shown in Figure 2b. The frequency tuning range is highly related to the local resonant mode dispersion and thermally-mediated mode-splitting frequency shift resulting from the optical power absorption in the microresonator. Mode-hopping instabilities are also observed but these can be suppressed by optimizing the microresonator operation temperature. In support of the measurements, the LLE numerical modeling of Figure 2c (detailed in Methods) verifies the comparative deterministic parametric oscillations when the mode frequency shifts are introduced at $\mu = 15$ and 37.

**Tunable radiation and multi-frequency terahertz oscillators generation.** With the designed cavity mode dispersion and differential TE-TM thermal dependence, the tunable OPO is facilitated by gradually changing the pump-resonance detuning, where the optical power absorption leads to increase (decrease) of the resonant mode temperature by forward (backward) pump wavelength tuning. The generated tunable OPO is passively down-converted to the terahertz radiation via a plasmonic nanoantenna array built on the InAs substrate [53]. The incident optical beam excites surface plasmons on the nanoantenna array, providing strong enhancement of the generated near-field photocarriers, subsequently accelerated by the surface built-in electric field in the InAs substrate without any external bias voltage. Since the nanoantenna array operates without any bias voltage, the direct current through the emitter is eliminated, leading to minimal heating. Consequently, the shot noise and thermal noise contributed from the nanoantenna array itself is believed to be negligible. Figure 3a illustrates the terahertz signal generation and detection schematic. By optimizing the intracavity power and detuning, terahertz radiation is generated over a broad frequency range (detailed in Methods and Supplementary Information II). To generate tunable terahertz radiation, two adjacent resonant modes ($\mu' = 9500$ and 9501, where $\mu'$ is the absolute cavity mode number) are examined and optically pumped, which is subsequently down-converted to radio frequency (RF) spectral ranges using a harmonic mixer to examine the signal-to-noise ratio (SNR), frequency tunability (intra- and inter-resonant modes), and frequency stability. The detected terahertz radiation spectra span from 330 to 750 GHz (bounded only by harmonic mixer frequency bandwidth) as shown in Figure 3b and feature an ≈ 50 dB SNR, apart from the 550 and 750 GHz frequencies where strong water vapor absorption occurs. The fluctuations in frequency and power are included in standard deviation error bars of Figure 3b as well. Subsequently, in Figure 3c, we examine the terahertz frequency tuning range over a single resonant mode ($\mu' = 9500$), with tunable optical parametric sidebands covering from 1586 to 1570 nm where the spectral gaps are related to local phase mismatching of the Turing instabilities. The spectral gaps require more accurate phase matching between the pump and signal by optimizing the dispersion engineering and operating temperature. The corresponding pump laser absolute frequencies are shown in Figure 3c inset over the thermally broadened cavity resonance. Power equalization between the pump and parametric sidebands are performed to optimize photomixing efficiency. The radiated terahertz power is measured with lock-in detection using a pyroelectric detector at a 20 Hz. Figure 3d shows the detected terahertz power across 2.8-octaves from 330 GHz to 2.3 THz, which decreases at the higher radiation frequencies and then rolls off to the noise floor of the pyroelectric detector. The detected power above 500 GHz agrees well with the simulated magnitude response of the nanoantenna array denoted with the dashed gray curve. Below 500 GHz, the power deviation arises from spectral filtering as well as degraded terahertz beam coupling inside the pyroelectric detector.

By optimizing the microresonator intracavity power and pump-resonance detuning, the sub-comb lines around the parametric sidebands can be locally excited. Figure 3e shows two example optical cluster frequency comb spectra with the signal-pump frequency spacing of 553 and 651 GHz. The generated frequency comb clusters are subsequently down-converted to the terahertz spectral range to generate multi-frequency terahertz oscillators with line frequency spacing of 19.86 GHz at the 553 and 651 GHz center frequencies. Figures 3f and 3g show the generated multi-frequency terahertz signal after down-conversion to the RF spectral range via changing the harmonic mixer reference frequency to down-convert all frequency tones simultaneously.

**Active feedback stabilization of the generated terahertz radiation.** Tunability of the generated single and multi-frequency terahertz wave highly depends on effective microresonator temperature which, in turn, introduces thermal stochastic frequency fluctuations of the generated terahertz signal induced by thermodynamical fluctuations of the microresonator mode. The small mode volume of the microresonator physically leads to high thermal sensitivity to ambient temperature fluctuations. The thermal instability originates from thermal energy exchange between the microresonator and ambient background or laser heating from the intracavity optical power absorption. The thermal fluctuations, fundamentally mediated by thermal expansion and thermo-refractive index effects, impose phase decoherence of the generated terahertz signal.

The microresonator mode optical frequency is denoted as $v_{\mu'} = \mu' f_{FSR} + f_0$ where $f_{FSR}$ is the mode free spectral range, and $f_0$ is the offset frequency. To improve the instantaneous linewidth and frequency stability of the generated terahertz signal and further increase its frequency tuning resolution, microresonator intracavity power stabilization (main stabilization loop) and the pump-resonance detuning stabilization (auxiliary stabilization loop) are engaged, as illustrated in Figure 4a. The frequency tuning resolution is fundamentally limited by the frequency fluctuations of the microresonator FSR, $\Delta f_{FSR}(T)$. With the microresonator thermodynamics [54], the frequency fluctuations variance behaves as $\langle \Delta f_{FSR}^2 \rangle = |\eta_{FSR}|^2 \frac{k_B T^2}{\rho C V}$ where $\eta_{FSR} = df_{FSR}/dT$ is the thermal transduction coefficient, $\rho$ is the material density, $C$ is the specific heat, and $V$ is the microresonator mode volume. Thermal transduction coefficient can be decomposed via $\eta_{FSR} = \frac{dv_{\mu'}}{dT}\left(\frac{\partial f_{FSR}}{\partial v_{\mu'}} + \frac{2}{\varepsilon}\frac{\partial f_{FSR}}{\partial \delta}\right)$, where $\varepsilon$ is the mode linewidth [55]. The measured temperature-dependent mode frequency change is $dv_{\mu'}/dT = 39.2\ GHz/K$ and $df_{FSR}/dv_{\mu'} \approx 1/\mu'$. To mitigate the thermal noise transduction, $\frac{\partial f_{FSR}}{\partial v_{\mu'}} + \frac{2}{\varepsilon}\frac{\partial f_{FSR}}{\partial \delta} = 0$ should be satisfied which requires the stabilization the intracavity power via controlling the pump laser power and frequency.

With the terahertz signal generation and active feedback stabilization setup (Figure 4a), the linearly polarized and power equalized optical beam is injected into the plasmonic nanoantenna array. The generated terahertz signal is down-converted to the RF domain by a harmonic mixer. The detected RF signal is managed by a series of electronics. An electrical spectrum analyzer and a signal source analyzer are used to characterize instantaneous linewidth and frequency noise of the down-converted signal at the intermediate frequency. The measured frequency noise of the down-converted signal is limited by the reference signal (LO1). To facilitate feedback locking, electrical frequency division method is utilized to further decrease the carrier frequency of the down-converted signal which subsequently is referenced to another source (LO2). The timing referenced frequency counter continuously samples the divided signal to examine the frequency stability of the terahertz signal at the free-running and stabilized conditions. The extracted error signal is managed by a proportional-integral-derivative (PID) controller to control the pump laser polarization to stabilize the microresonator intracavity power after polarization demodulation with a linear polarizer. Meanwhile, the integral error signal auxiliary controls pump laser frequency to equivalently stabilize the pump-resonance detuning ($\delta$). We realize a dual-step stabilization protocol including pump laser power locking followed by pump laser frequency locking. Figure 4b shows the free-running and resultant stabilized instantaneous linewidth of the generated terahertz signal. After pump laser power and frequency stabilization, the Gaussian-fitted linewidth is improved ≈ 105,000× for a 39-ms sweep time.

After terahertz signal frequency stabilization, the frequency noise power spectral density (PSD) is measured as shown in Figure 4c along with the frequency noise of the reference source (LO1), the $\beta$ separation line, and the flicker frequency noise power law (1/$f$) where the 100-Hz linewidth is observed. The low Fourier frequency flicker noise originates from the pump laser amplifier intensity noise conversion [56]. The instantaneous linewidth and frequency noise PSD characterize the short-term frequency stability. The long-term frequency stability is examined with a referenced frequency counter for a 300-ms gate time (detailed in Methods and Supplementary Information III). Figure 4d shows the Allan deviation of the free-running signal and residual signal at 651.5 GHz. The Allan deviation of the reference signal (LO2) at 4.44841 MHz is included as well. The residual frequency stability is improved by 400,000× for a 1-second averaging time and reaches the instrumental limit of the frequency counter. The stabilized terahertz signal has a frequency tuning range and resolution of 14 kHz and 27 ± 1.5 Hz, respectively (Figure 4e). The stabilized frequency tuning range is dependent on the feedback locking bandwidth. The frequency tuning resolution is determined by the frequency tuning step of the reference source 2 and bounded by the reference source linewidth and the residual noise of the active feedback locking loop.

**Discussion.** We report a chip-scale coherent frequency-agile terahertz radiation synthesizer operating at room temperature based on an integrated microresonator parametric oscillator and a plasmonic nanoantenna array, capable of providing both single-frequency terahertz radiation as well as multi-frequency terahertz oscillators in the same system platform. Through optical-to-terahertz frequency division and without requiring optical spectral shaping, the achieved 2.8-octave frequency tunability is, to our best knowledge, the largest octave range realized in chip-scale semiconductor devices with a single pump laser. Moreover, the ≈ 27 Hz frequency tuning resolution for the 651 GHz carrier frequency is one of the highest resolution achieved to date. With the deterministic pump-resonance laser detuning and hybridized modes with thermal control, we report less than 100 Hz linewidths of the terahertz radiation with feedback stabilization of the microresonator intracavity power. The dual pump power- and frequency-stabilization improves the long-term frequency stability by five orders-of-magnitude, reaching an instrument-limited relative frequency residual instability of 1×10$^{-15}$ at the averaging time of 1,000 seconds. With our nonlinear parametric oscillation and plasmonic array photomixing, the high-spectral-purity frequency-agile terahertz synthesizer in single-frequency, tunable and multi-frequency realizations enable spectroscopy and waveform synthesis in the terahertz regime, supporting advances in metrology, sensing, frequency-agile imaging and wireless communications.

**Methods**

**Integrated dispersion-managed microresonator fabrication:** A 3 μm thick SiO$_2$ layer is deposited via plasma-enhanced chemical vapor deposition (PECVD) on a *p*-type 8" silicon wafer serving as an under-cladding. Then an 800 nm nitride layer is deposited via low-pressure chemical vapor deposition and patterned by the optimized 248 nm deep-ultraviolet lithography and etched down to the buried oxide cladding via optimized reactive ion dry etching. The etched sidewalls have an etch verticality of 88° characterized by a scanning electron microscope. The nitride rings are then over-cladded with a 4.5 μm thick oxide layer.

**Microresonator group velocity dispersion and avoided-mode-crossing characterization**: A tunable laser (Santec, TSL-510) is swept over a wavelength range from 1520 to 1610 nm at a sweeping speed of 30 nm/s with an output power of 2 mW to measure the cold cavity mode spectra of the microresonator. The 1% of the laser output power is injected into a fiber coupled hydrogen cyanide gas cell (HCN-13-100, Wavelength References Inc.) to calibrate the cavity mode frequency for extracting the cavity mode group velocity dispersion and wavelength-dependent cavity loaded quality factor. The transmissions of the microresonator and the gas cell are recorded during the laser sweep by a data acquisition system with sampling frequency of 1.7 MHz determined by an unbalanced fiber Mach-Zehnder interferometer (MZI).

**Simulated parametric oscillation with Lugiato-Lefever equation:** Taking the normal group velocity dispersion (GVD) and avoided mode-crossing into consideration, the tunable optical parametric oscillation dynamics are numerically modeled with the Lugiato-Lefever equation (LLE) written as:

$$t_R \frac{\partial E(t,\tau)}{\partial T} + i \left( \frac{\beta_2}{2} \frac{\partial^2}{\partial \tau^2} - \gamma |E|^2 \right) E(t,\tau) + (\alpha + i\delta) E(t,\tau) = i\sqrt{T} E_{in} \qquad (1)$$

where $t_R$ = 50.3 ps, $\beta_2$ = 18.31 fs$^2$/mm, $\gamma$ = 1 W$^{-1}$m$^{-1}$, $Q$ = 1.55×10$^6$, $\alpha$ = 0.018, and $T$ = 0.009. The pump power $P_{in}$ and wavelength $\lambda_p$ are set at 1.4 W and 1588 nm. We numerically solve the equation with split-step Fourier method initiating from quantum noise. The tunable parametric oscillation can be excited at the specific microresonator modes by properly setting the avoided-mode frequency shift.

**Tunable terahertz wave and multi-frequency terahertz oscillators generation:** The generated parametric oscillation with an average power of ≈ 20 dBm is injected into an optical preamplifier after an optical bandpass filter. The amplified optical beam is focused by a biconvex lens with a focus length of 10 cm into the plasmonic nanoantenna array. The generated terahertz signal is collimated and focused by a pair of gold-coated parabolic mirrors. The signal is then received by a harmonic mixer (VDI MixAMC with WR 1.5 and WR 2.2 horn antennas) to down-convert into the RF domain for linewidth and frequency stability examination or by a pyroelectric detector (QMC Instruments Ltd) with a lock-in amplifier (SR830) for terahertz power measurements.

**Active stabilization of the terahertz signal:** The generated terahertz signal is down-converted to an intermediate frequency (IF) of 1.154 GHz with the harmonic mixer (VDI MixAMC with WR 2.2 horn antenna) referenced to a local oscillator (LO1, Gigatronics 905) at a frequency of LO1 = 12.043 GHz. The IF signal is then amplified by an electrical amplifier (Mini-circuits, ZVA-443HGX+) to ≈ -10 dBm and monitored by an electrical spectrum analyzer (ESA, Agilent MXA9020A). The amplified signal is frequency divided with a 260× RF frequency divider (RF bay, FPS-260-4) to the frequency of 4.44 MHz, which is subsequently filtered out by a low-pass filter (LPF, Mini-circuits, BLP-5+). The divided signal is referenced to the local oscillator 2 (LO2) to obtain phase error signal. The error signal is filtered by another low pass filter (LPF) with a cutoff frequency of 1 kHz and routed to a proportional-integral-differential controller (Vescent photonics, DS-125) to generate feedback signal. The feedback signal controls the pump laser polarization via an electrical-controlled polarization controller (EPC, EOSPACE) combined with a polarization beam splitter to stabilize the intracavity power. Furthermore, the auxiliary of the PID controller is used to control pump laser frequency to equivalently stabilize the pump-resonance detuning.

## Data Availability

All the data and methods are present in the main text and the supplementary materials. The raw datasets generated during and/or analysed during the current study are available from the corresponding author upon request.

**Code Availability**

Codes used in this work are available from the authors upon request.

34. T. J. Kippenberg, A. L. Gaeta, M. Lipson, and M. L. Gorodetsky, Dissipative Kerr solitons in optical microresonators. *Science* **361**, 8083 (2018).

35. B. C. Yao, S. W. Huang, Y. Liu, A. K. Vinod, C. Choi, M. Hoff, Y. N. Li, M. B. Yu, Z. Y. Feng, D. L. Kwong, Y. Huang, Y. J. Rao, X. F. Duan, and C. W. Wong, Gate-tunable frequency combs in graphene-nitride microresonators. *Nature* **558**, 410-415 (2018).

36. S. W. Huang, H. Zhou, J. H. Yang, J. F. McMillan, A. Matsko, M. Yu, D. L. Kwong, L. Maleki, and C. W. Wong, Mode-locked ultrashort pulse generation from on-chip normal dispersion microresonators. *Phys. Rev. Lett.* **114**, 053901 (2015).

37. M. G. Suh, Q.-F. Yang, K. Y. Yang, X. Yi, and K. J. Vahala, Microresonator soliton dual-comb spectroscopy. *Science* **354**, 600-603 (2016).

38. P. Marin-Palomo, J. N. Kemal, M. Karpov, A. Kordts, J. Pfeifle, M. H. P. Pfeiffer, P. Trocha, S. Wolf, V. Brasch, M. H. Anderson, R. Rosenberger, K. Vijayan, W. Freude, T. J. Kippenburg, and C. Koos, Microresonator-based solitons for massively parallel coherent optical communications. *Nature* **546**, 274-279 (2017).

39. J. Feldmann, N. Youngblood, M. Karpov, H. Gehring, X. Li, M. Stappers, M. Le Gallo, X. Fu, A. Lukashchuk, A. S. Raja, J. Liu, C. D. Wright, A. Sebastian, T. J. Kippenberg, W. H. P. Pernice, and H. Bhaskaran, Parallel convolutional processing using an integrated photonic tensor core. *Nature* **589**, 52-58 (2021).

40. X. Y. Xu, M. X. Tan, B. Corcoran, J. Y. Wu, A. Boes, T. G. Nguyen, S. T. Chu, B. E. Little, D. G. Hicks, R. Morandotti, A. Mitchell, and D. J. Moss, 11 TOPS photonic convolutional accelerator for optical neural networks. *Nature* **589**, 44-51 (2021).

41. M. G. Suh and K. J. Vahala, Soliton microcomb range measurement. *Science* **359**, 884-887 (2018).

42. P. Trocha, M. Karpov, D. Ganin, M. H. P. Pfeiffer, A. Kordts, S. Wolf, J. Krockenberger, P. M. Palomo, C. Weimann, S. Randel, W. Freude, T. J. Kippenberg, and C. Koos, Ultrafast optical ranging using microresonator soliton frequency combs. *Science* **359**, 887-891 (2018).

43. J. Riemensberger, A. Lukashchuk, M. Karpov, W. L. Weng, E. Lucas, J. Q. Liu and T. J. Kippenberg, Massively parallel coherent laser ranging using a soliton microcomb. *Nature* **581**, 164-170 (2020).
16

**Acknowledgements**

The authors acknowledge fruitful discussions with Dr. Yoonsoo Jang, Prof. Ming Xin, Jaime Gonzalo Flor Flores, Dr. Jinkang Lim, Prof. Heng Zhou, Prof. Shu-Wei Huang, Dr. Jinghui Yang, and Futai Hu on simulations and general notes. Wong's group acknowledges financial support from the Lawrence Livermore National Laboratory (B622827), the National Science Foundation (1824568, 1810506, 1741707, 1829071), and the Office of Naval Research (N00014-14-1-0041). Jarrahi's group acknowledges financial support from the Office of Naval Research (N00014-19-1-2052). The development of the plasmonic nanoantenna arrays used for photomixing is supported by the Department of Energy (# DE-SC0016925).

**Author contributions**

W.T.W., M.J., and C.W.W. initiated the project. W.T.W., and P. K. L. conducted the experiments. W.T.W., and P. K. L. analyzed the data and performed the simulations. A. K., D. T., J. M., and H. L. contributed to the experiments. M.Y. and D.-L.K. performed the device nanofabrication. W.W., P. K. L., M.J., and C.W.W. contributed to writing and revision of the manuscript.

**Competing interests**

The authors declare no competing financial or non-financial interests.

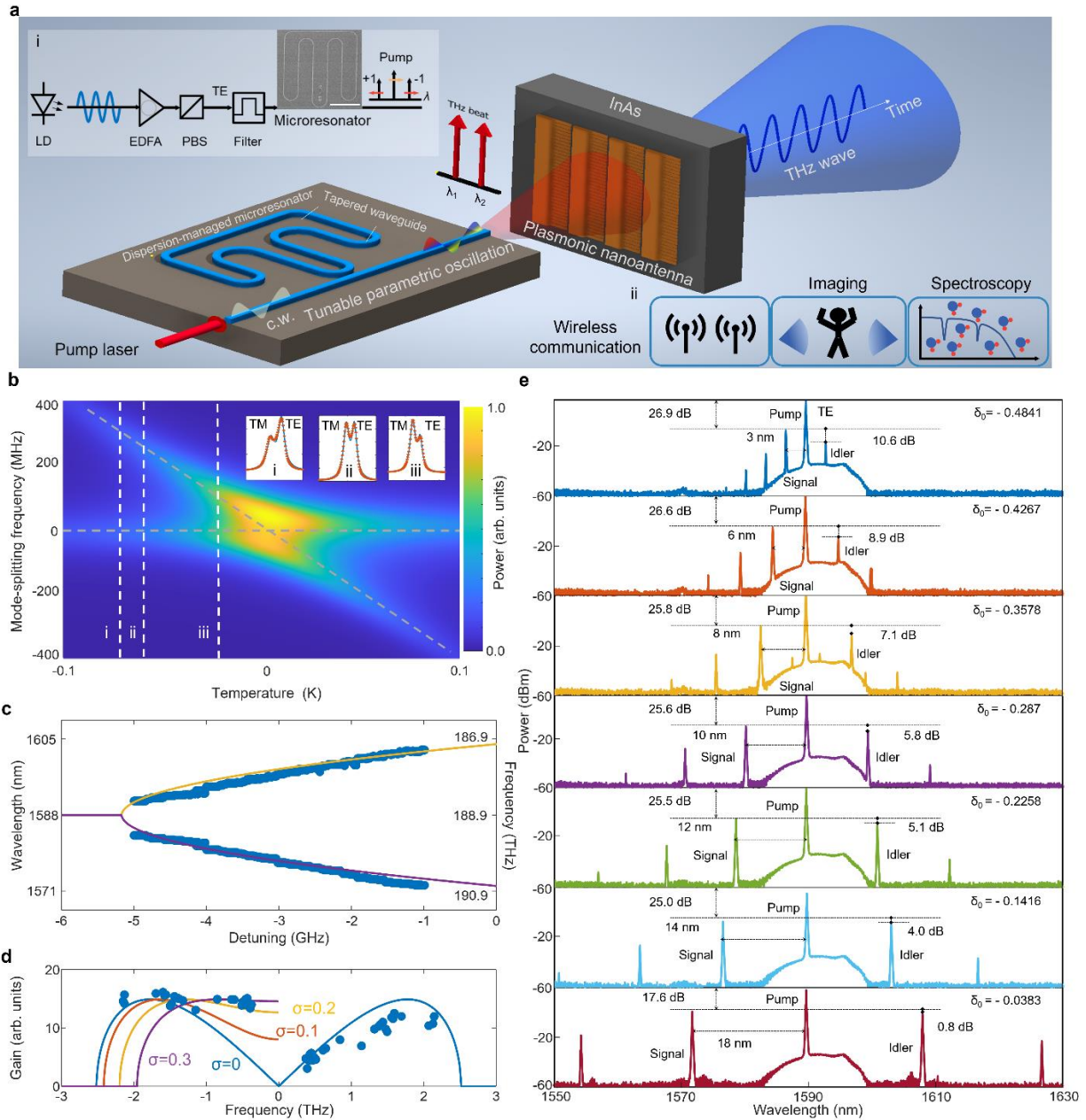

**Figure 1 | Chip-scale THz radiation source based on the high-*Q* dispersion-managed nitride parametric microresonator and plasmonic photoconductive InAs nanoantenna array photomixing. a,** Illustration of the coherent 2.8-octave tunable terahertz parametric radiation. Inset **i**: Tunable optical parametric oscillation generation process with microresonator. LD: laser diode, EDFA: erbium-doped fiber amplifier, PBS: polarization beam splitter, TE: transverse electric, c.w.: continuous wave. Scanning electron micrograph with scale bar of 400 μm. Inset **ii**: Applications of the broadly-tunable coherent terahertz parametric radiation. **b,** Modeled hybridized mode

spectra evolution from the TE-TM coupled modes with swept cavity temperatures, TM: transverse magnetic. Inset **i, ii, iii**: Measured hybridized mode spectra at different mode temperatures measured with swept-wavelength interferometry and fitted by the double-Lorentzian model. **c,** Generated broadly-tunable parametric oscillation via continuously swept pump-resonance detuning. **d,** Parametric gain oscillation with and without coupled-mode frequency shift, matched with the experimentally measured parametric oscillation peak power. **e,** Representative microresonator tunable parametric oscillation under controlled pump-resonance detuning.

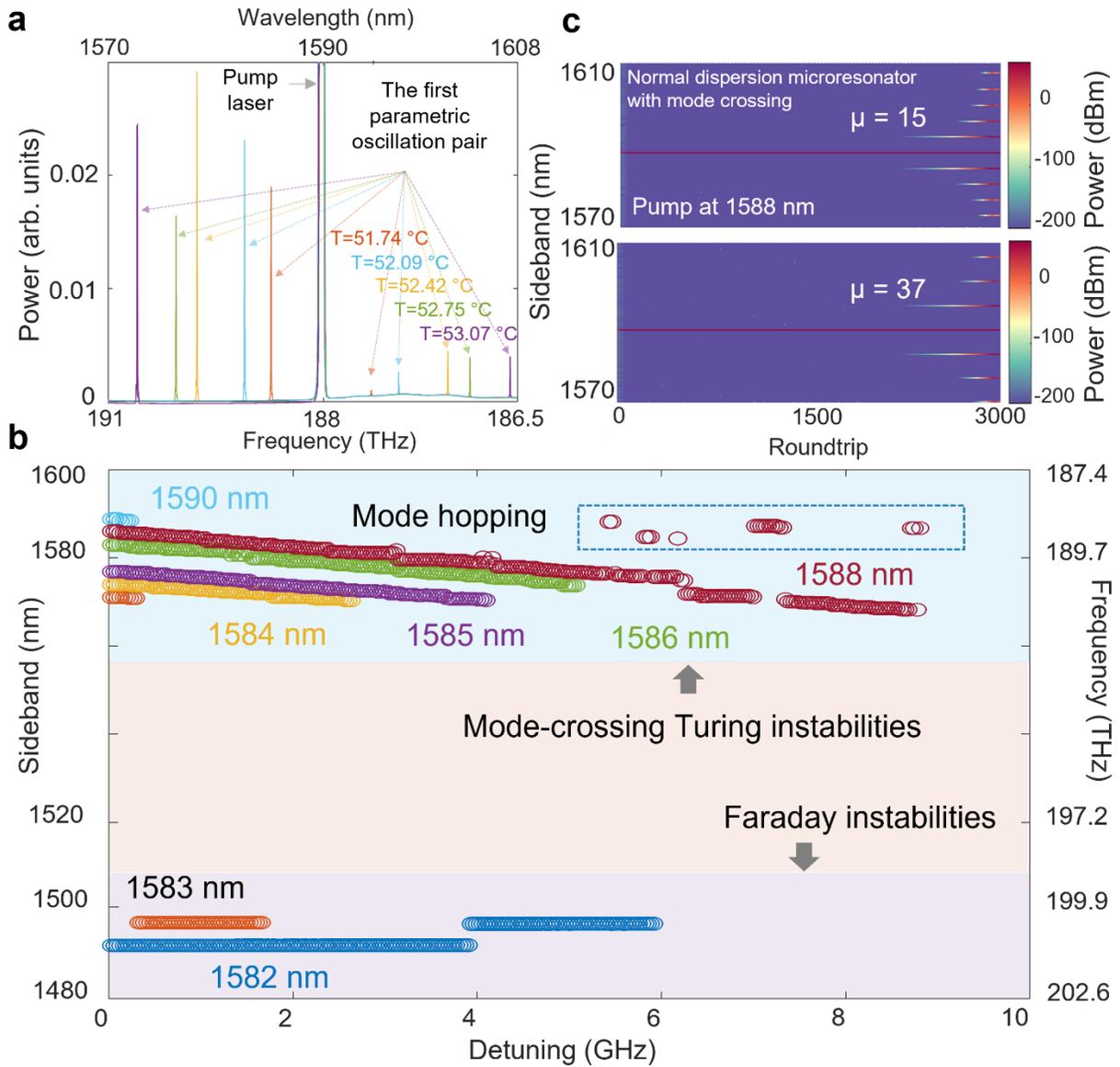

**Figure 2 | Broadly-tunable optical parametric oscillation in the dispersion-managed microresonators. a,** Measured parametric oscillation pair at different chip temperatures from 51.74 to 53.07 °C while maintaining constant pump power and optimal pump detuning to obtain the initial oscillation pair. **b**, Pump wavelength dependence of parametric oscillation. Competition between different parametric oscillations, determined by the intrinsic Faraday ripple [51] and thermal-related mode frequency shift, is observed. **c**, Lugiato-Lefever modelling with avoided-mode-crossings for parametric sidebands of 0.3 and 0.74 THz tunability. $\mu$ is the azimuthal mode number with respect to the pump cavity mode.

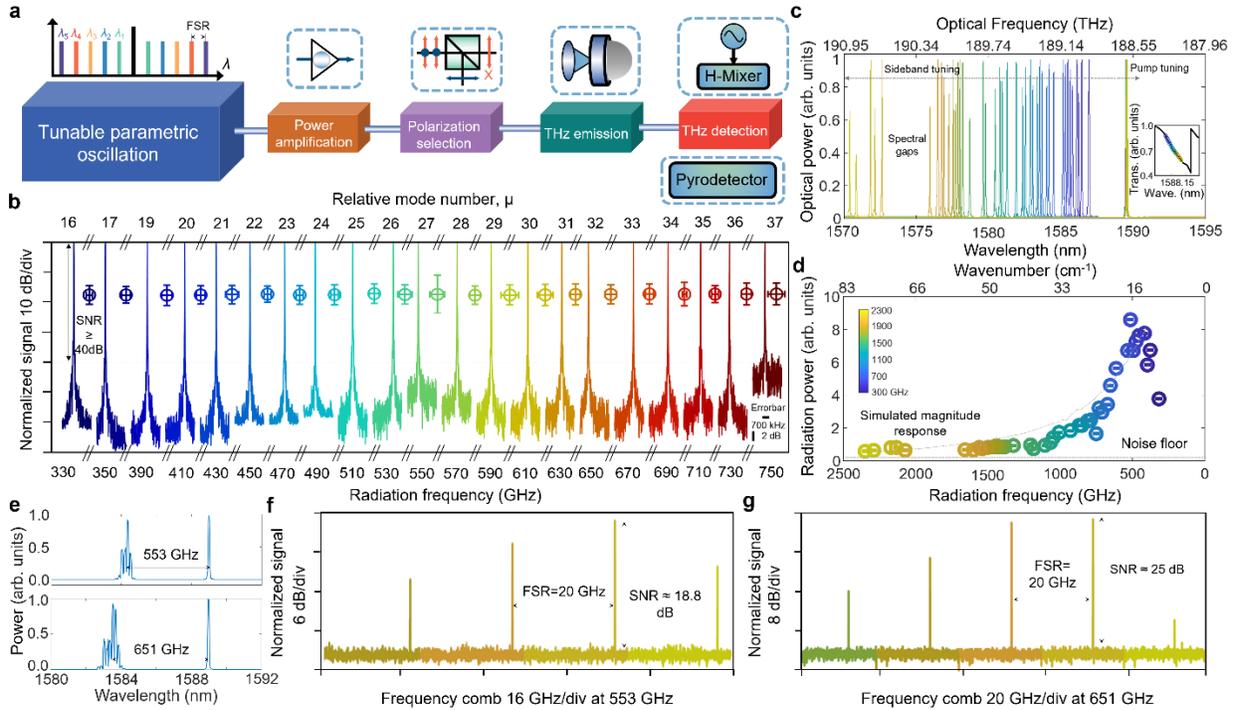

**Figure 3 | 2.8-octave tunable radiation and multi-frequency terahertz oscillators through photomixed microresonator optical parametric oscillation. a**, Schematic of the terahertz generation and detection setup, FSR: Free-spectral range. **b**, Measured down-converted RF spectra showing ≈ 100 kHz 3-dB instantaneous linewidth and more than 50 dB signal-to-noise ratio (SNR) (apart from the 550 and 750 GHz terahertz radiation frequencies), corresponding to generated radiation from 330 to 750 GHz. Experimental error bars for each carrier are illustrated too. **c,** Measured tunable parametric sidebands via continuously smaller pump-resonance detuning in one cavity resonance ($\mu' = 9,500$). Inset: the hot cavity resonance including Kerr and thermally-induced resonance red shift (≈ 14.68 GHz) and the corresponding color-coded pump-resonance detuning. **d,** Measured 2.8-octave coherent radiation power from 330 GHz to 2.3 THz with a pyroelectric detector and lock-in amplification. **e,** Measured optical spectra with sub-comb lines when optimizing the microresonator intracavity power. **f** and **g,** Measured multi-frequency terahertz oscillators at center frequency of 553 and 651 GHz corresponding the top and bottom panel of **e**.

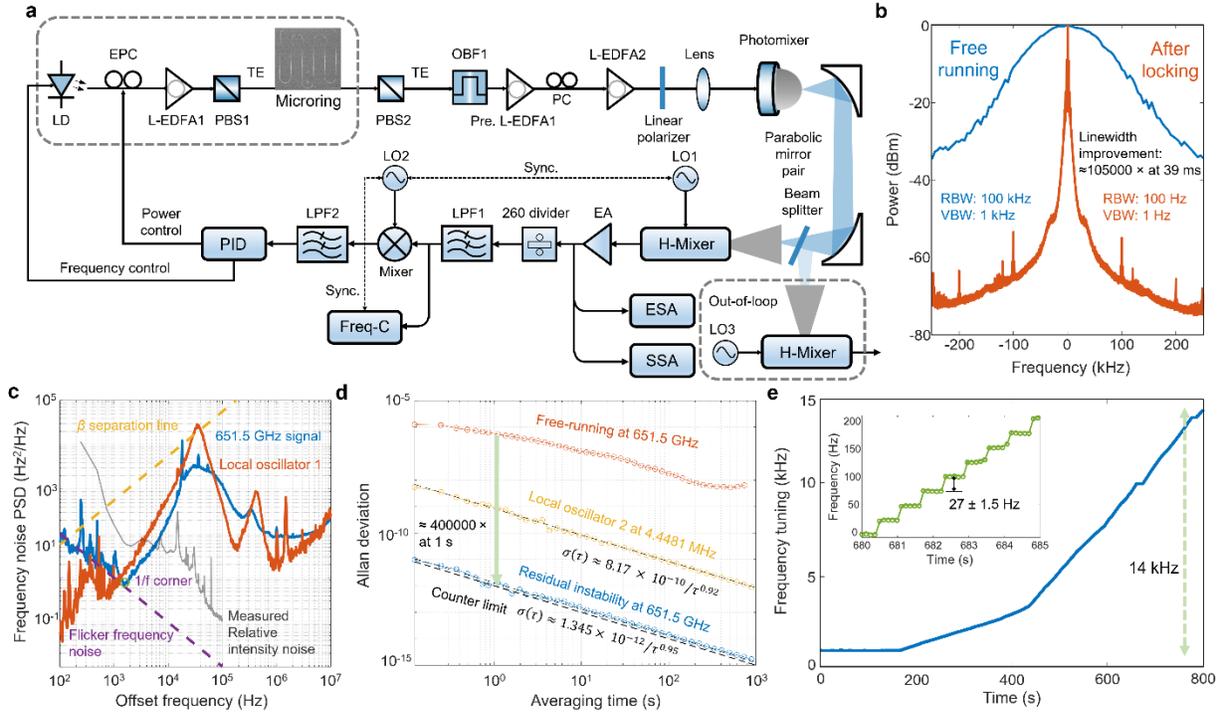

**Figure 4 | Sub-100-Hz frequency stabilization of the terahertz radiation via feedback control of the microresonator intracavity power. a**, Terahertz generation and feedback control experimental setup. EPC: electrically controlled polarization controller, OBF: optical bandpass filter, H-Mixer: harmonic mixer, LO: local oscillator, EA: electrical amplifier, ESA: electrical spectrum analyzer, SSA: signal source analyzer, LPF: low pass filter, Freq-C: frequency counter, PID: proportional-integral-derivative controller. **b**, Measured RF beatnote spectra of the free-running and stabilized signals, showing a 105,000× improvement of the 3-dB Gaussian linewidth after frequency stabilization at a sweeping time of 39 ms, RBW: Resolution bandwidth; VBW: Vadio bandwidth. **c**, Frequency noise power spectral density (PSD) of the RF beatnote down-converted from a frequency-stabilized 651.5 GHz radiation, along with the PSD of local oscillator 1 (LO1), the $\beta$-separation line, measured relative laser intensity noise, and flicker frequency noise curve. **d**, Allan deviation of the free-running and stabilized terahertz signal along with Allan deviation of the referenced local oscillator (LO2). The stabilized signal shows the relative frequency deviation of $\sigma(\tau) = 1.345 \times 10^{-12}/\tau^{0.95}$. **e**, Deterministically tuned terahertz radiation after frequency stabilization over a 14 kHz range, with ≈ 27 ± 1.5 Hz discrete tuning steps as shown in inset and only by LO2 frequency resolution.

23